\documentclass{ws-procs9x6-cpt19}
\usepackage{accents}
\usepackage{color}
\usepackage{xcolor}
\usepackage{amsmath}
\usepackage{gensymb}
\usepackage[utf8]{inputenc}

\def\lsim{\mathrel{\rlap{\lower4pt\hbox{\hskip1pt$\sim$}}
    \raise1pt\hbox{$<$}}}
\def\gsim{\mathrel{\rlap{\lower4pt\hbox{\hskip1pt$\sim$}}
    \raise1pt\hbox{$>$}}}

\newcommand{\beq}{\begin{eqnarray}}
\newcommand{\eeq}{\end{eqnarray}}

\def\ismean{\accentset{\circ}{a}^{(3)}} 
\def\ismecn{\accentset{\circ}{c}^{(4)}} 
\def\ismetn{\accentset{\circ}{a}^{(5)}}
\def\ismegn{\accentset{\circ}{c}^{(6)}}

\begin{document}

\title{Search for Lorentz Violation Using High-Energy Atmospheric Neutrinos In IceCube}

\author{
  Carlos~A.~Arg\"{u}elles$^1$
}

\address{
  $^1$Massachusetts Institute of Technology, Cambridge, MA 02139, USA
}

\author{On behalf of the IceCube Collaboration}

\begin{abstract}
 High-energy atmospheric neutrinos observed by the IceCube Neutrino Observatory are extremely sensitive probes of Lorentz violation (LV). Here we report the result of analyzing two years of IceCube data in the search for LV. This analysis places some of the strongest constraints on LV when considering high-dimensional operators.
\end{abstract}

\bodymatter

\section{Lorentz violation effects on high-energy neutrinos}

Neutrino flavor information is the one of the most sensitive observables to Lorentz symmetry violation (LV).
This is due to the fact that neutrino flavor changes are produced by the interference between different neutrino states.
It is only through the study of neutrino flavor morphing that we have been able to infer the existence of neutrino masses; direct neutrino mass measurements have so far yielded null results~\cite{Tanabashi:2018oca}.
The power of neutrino interferometry has been demonstrated in the study of neutrino and anti-neutrino oscillation parameters, which provides a strong test for CPT symmetry~\cite{Barenboim:2017ewj,Barenboim:2017vlc}.
The neutrino to anti-neutrino mass-squared differences are limited to $\left|\Delta m_{31}^{2}-\Delta \overline{m}_{31}^{2}\right|<0.8 \times 10^{-3}~\mathrm{eV}^{2}$ while the neutral kaon mass difference~\cite{Barenboim:2017ewj} is only bounded to be $\left|m^{2}\left(K^{0}\right)-m^{2}\left(\overline{K}^{0}\right)\right|<0.25 ~\mathrm{eV}^{2}$.

\begin{figure}[h!]
    \centering
    \includegraphics[width=0.55\textwidth]{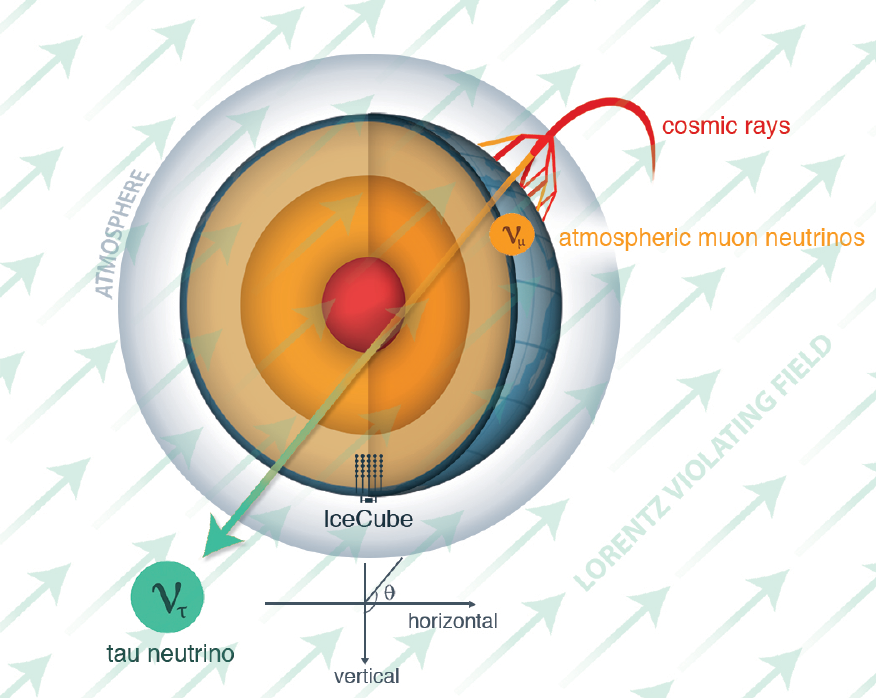}~\includegraphics[width=0.55\textwidth]{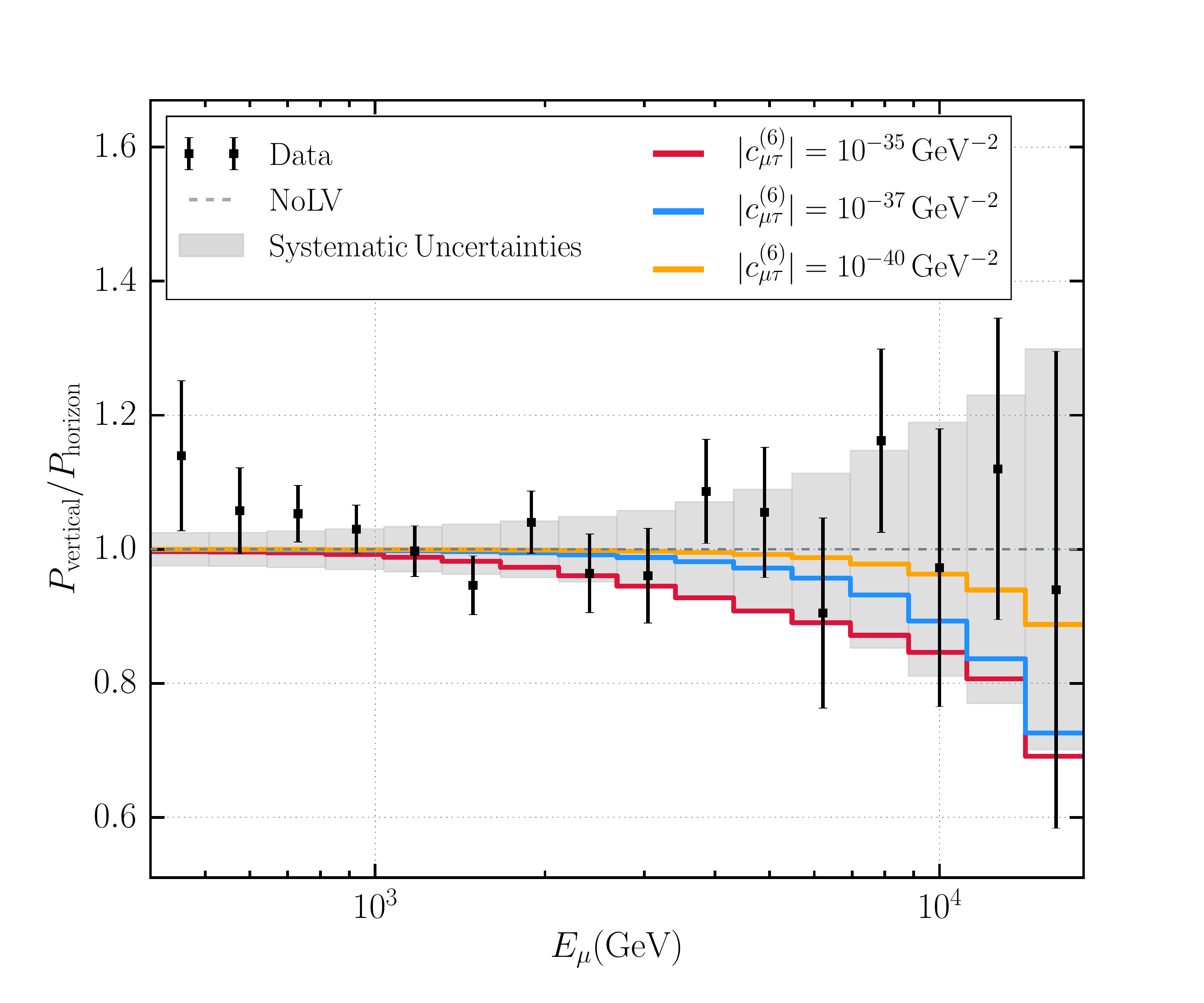}
    \caption{{\it \bf Illustration of the effect of Lorentz violation in high-energy atmospheric neutrinos.}  Left: Cartoon of the analysis, a muon-neutrino produced in a cosmic-ray air shower travels through the Earth and converts to a tau-neutrino due to interactions with the all-permeating LV field (green background arrows). Right: The double-ratio of vertical to horizontal events as a function of the reconstructed muon energy. A horizontal line at one corresponds to the Standard Model, other lines show the expected double ratio for various values of the Lorentz dimension six coefficient. Data points with statistical error bars are shown as well. The grey band is the total uncorrelated systematic uncertainty.}
    \label{fig:analysis_idea}
\end{figure}

We can incorporate the effect of LV into neutrino oscillations by considering the following Hamiltonian~\cite{Kostelecky:2011gq}
\beq
H\sim\frac{m^2}{2E}+\ismean-E\cdot\ismecn+E^2\cdot\ismetn-E^3\cdot\ismegn\cdots ,
\label{eq:Heff}
\eeq
where the first term is responsible for the measured neutrino oscillations and the latter terms encode the effect of LV.
The $a$-terms are CPT-odd, $c$-terms are CPT-even, and each index denotes the dimension of the operator that produced this term.
Each of these coefficients is a matrix with flavor indices.
In this work, we restrict ourselves to study the $\mu-\tau$ sector; see Fig.~\ref{fig:analysis_idea} (left).
At high energies, where $\frac{m^2}{2E}$ is negligible, the $\nu_\tau$ appearance probability is given by~\cite{GonzalezGarcia:2005xw}
\beq
P\left(\nu_{\mu} \rightarrow \nu_{\tau}\right) \sim \frac{\left|\begin{array}{l}{\accentset{\circ}{a}_{\mu \tau}^{(d)}-\accentset{\circ}{c}_{\mu \tau}^{(d)}}\end{array}\right|^{2}}{\rho_{d}^{2}} \sin ^{2}\left(\rho_{d} L \cdot E^{d-3}\right),
\label{eq:osc_prob}
\eeq
where we have introduced the LV strength
\beq
\rho_{d} \equiv \sqrt{\left(
\accentset{\circ}{a}_{\mu \mu}^{(d)
}\right)^{2}+\operatorname{Re}\left(
\accentset{\circ}{a}_{\mu\tau}^{(d)
}\right)^{2}+\operatorname{Im}\left(
\accentset{\circ}{a}_{\mu \tau}^{(d)
}\right)^{2}};
\eeq
similarly for $c$-terms.
As can be seen in Eq.~\eqref{eq:osc_prob}, the appearance probability increases with neutrino energy and travelled distance.
In this work we focus on the effect of LV on neutrino flavors, but it is worthwhile to note that other LV effects can be studied with atmospheric neutrinos.
LV can also affect the production of neutrinos in cosmic-ray air showers~\cite{Altschul:2013yja,Diaz:2016dpk}, but use of these effects to probe LV is expected to be less sensitive than neutrino interferometry.

\section{IceCube and high-energy atmospheric neutrinos}

The IceCube Neutrino Observatory is a gigaton-scale neutrino detector located in Antarctica~\cite{Aartsen:2016nxy}.
IceCube currently has the largest sample of neutrinos above a TeV due to its large effective area. These neutrinos are produced by cosmic-rays interacting in the Earth's atmosphere. At TeV energies, atmospheric neutrinos are primarily produced by kaon decay. At higher energies, neutrinos from charmed hadrons are expected to be important, but this component has yet to be observed. Finally, for energies above $\sim 100~ {\rm TeV}$, extraterrestrial neutrinos dominate the flux, and are expected to be the most sensitive to LV~\cite{Arguelles:2015dca}. Since we are looking for a flavor changing effect, we restrict the analysis to reconstructed muon energies below 18~TeV.
In this regime atmospheric neutrinos, whose flavor composition is well predicted, dominate the flux. The effect of the LV dimension six operator is illustrated in Fig.~\ref{fig:analysis_idea} (right). This shows the double ratio of horizontal-to-vertical data-to-prediction as a function of the muon energy. As the LV flavor-violating coefficient is increased muon-neutrinos start disappearing at the highest energies. 

\section{Results}

\begin{figure}
    \centering
    \includegraphics[width=0.55\textwidth]{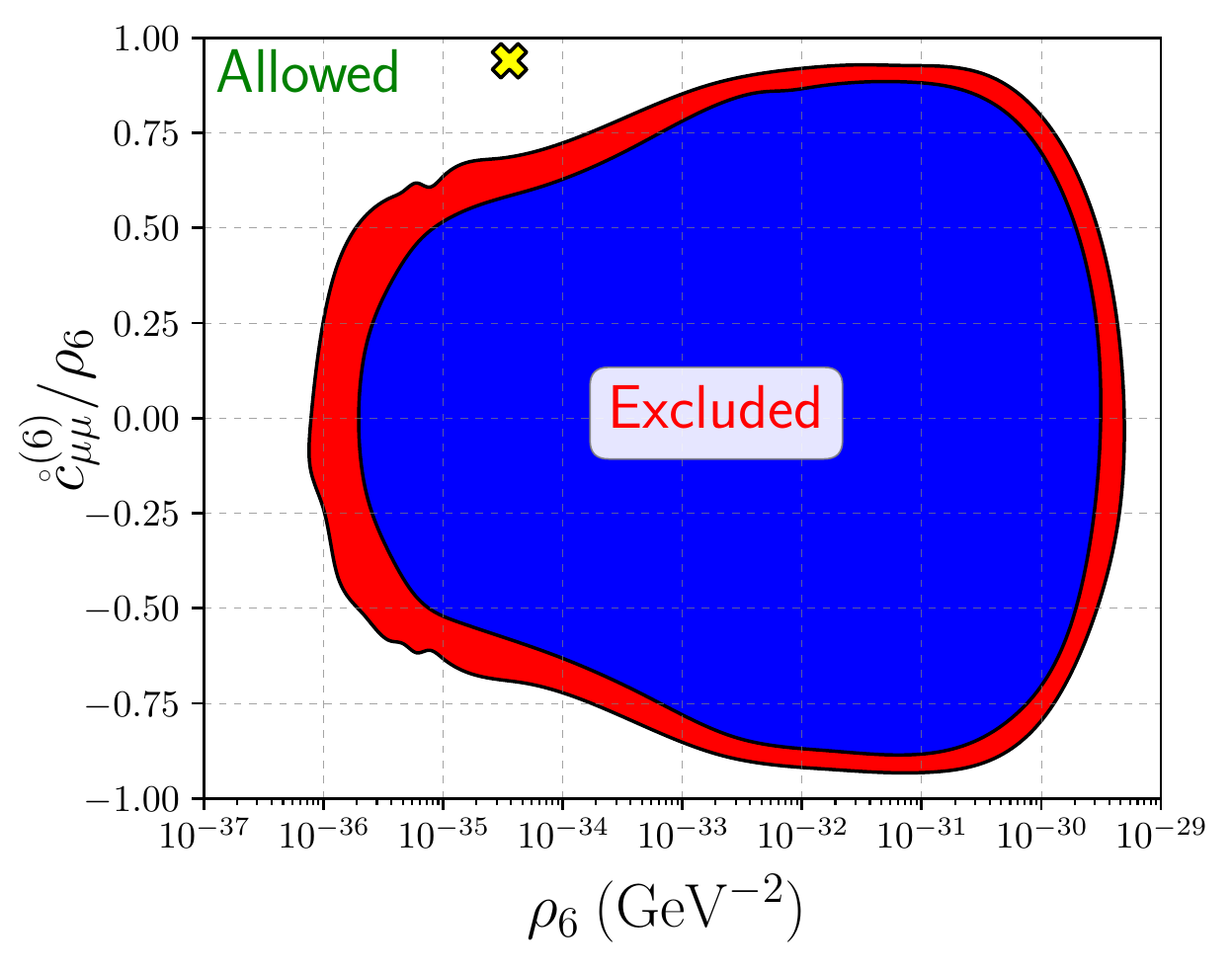}~\includegraphics[width=0.5\textwidth]{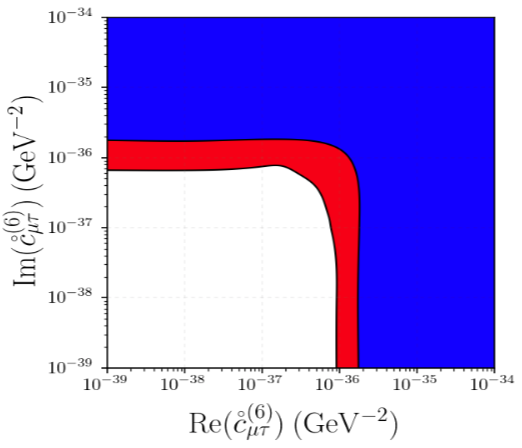}
    \caption{{\it \bf Constraints on Lorentz violation dimension six operator.}  Left: Constraints in term of the LV strength (horizontal axis) and the ratio of diagonal to total strength (vertical axis). The best-fit point of the analysis is marked by the standard yellow cross. Right: Constraints assuming a maximally violating flavor texture, {\it i.e.} diagonal elements are set to zero, as a function of the real and imaginary parts of the off-diagonal element. In both panels points inside the red (blue) patch are excluded at 90\% (99\%) C.L..}
    \label{fig:results}
\end{figure}

We have performed a search for muon-neutrino disappearance induced by Lorentz symmetry violation using two years of IceCube data described in~\cite{Aartsen:2015rwa}. This data set corresponds to through-going muon and was previously used to search for evidence of a high-energy astrophysical neutrino component.  This set contains 34975 events with a 0.1\% atmospheric muon contamination spanning an energy range from 400~GeV to 20~TeV muon energy. Using this data, we find no evidence for anomalous disappearance and place bounds on the flavor-violating coupling of neutrinos defined in the Lorentz symmetry violating standard model extension. The obtained constraints are among the strongest bounds of LV and are particularly important for higher dimensional operators~\cite{Aartsen:2017ibm}.
In Fig.~\ref{fig:results}, the results for the dimension six operator are shown. In the left panel it is shown as a function of the LV strength and the ratio of diagonal component size to LV strength. When the diagonal component dominates, no flavor change is introduced and thus no constraint is obtained. The constraint is strongest when the diagonal component vanishes, which is the case for maximal flavor violation. This limits the lower and upper parts of the constrained region. The smallest strengths are limited by the statistical uncertainty of high-energy atmospheric neutrinos and the largest strengths are limited by the uncertainty on the absolute normalization of the atmospheric flux. To compare with other results in the literature~\cite{Abe:2014wla}, we report our results for the maximum flavor violation scenario in Fig.~\ref{fig:results} (right). Finally, even though these results have been shown in terms of interactions between neutrinos and a hypothetical LV field, they can be recast to other scenarios such as neutrino-dark matter interactions~\cite{Capozzi:2018bps,Farzan:2018pnk,Klop:2017dim}.

\end{document}